\documentclass[conference]{IEEEtran}
\IEEEoverridecommandlockouts
\usepackage{cite}
\usepackage{amsmath,amssymb,amsfonts}
\usepackage{algorithmic}
\usepackage{graphicx}
\usepackage{textcomp}
\usepackage[table]{xcolor}
\usepackage{booktabs}
\usepackage[]{adjustbox}
\usepackage{subfig}
\usepackage{adjustbox}
\usepackage{graphbox}
\usepackage{multirow}
\usepackage{hyperref}
\usepackage{tikz}

\definecolor{ao(english)}{rgb}{0.0, 0.5, 0.0}
\definecolor{dandelion}{rgb}{0.94, 0.88, 0.19}
\definecolor{flax}{rgb}{0.93, 0.86, 0.51}
\definecolor{maize}{rgb}{0.98, 0.93, 0.37}
\def\BibTeX{{\rm B\kern-.05em{\sc i\kern-.025em b}\kern-.08em
		T\kern-.1667em\lower.7ex\hbox{E}\kern-.125emX}}	
	
\newcommand{\low}{\tikz\draw[red,fill=red] (0,0) circle (.5ex);}
\newcommand{\med}{\tikz\draw[maize,fill=maize] (0,0) circle (.5ex);}
\newcommand{\hig}{\tikz\draw[ao(english),fill=ao(english)] (0,0) circle (.5ex);}
\newcommand{\dc}{\tikz\draw[gray,fill=gray] (0,0) circle (.5ex);}

\begin{document}
\title{Interpreting Contrastive Embeddings in Specific Domains with Fuzzy Rules}

\author{
	\IEEEauthorblockN{Javier Fumanal-Idocin, Mohammadreza Jamalifard, Javier Andreu-Perez}
		\IEEEauthorblockA{\textit{School of Computer Science and Electronic Engineering}\\
			 \textit{University of Essex}\\ 
			 Colchester, United Kingdom \\
			 \{j.fumanal-idocin, jm23525, j.andreu-perez\}@essex.ac.uk}
	}
\maketitle

\begin{abstract}
Free-style text is still one of the common ways in which data is registered in real environments, like legal procedures and medical records. Because of that, there have been significant efforts in the area of natural language processing to convert these texts into a structured format, which standard machine learning methods can then exploit. One of the most popular methods to embed text into a vectorial representation is the Contrastive Language-Image Pre-training model (CLIP), which was trained using both image and text. Although the representations computed by CLIP have been very successful in zero-show and few-shot learning problems, they still have problems when applied to a particular domain. In this work, we use a fuzzy rule-based classification system along with some standard text procedure techniques to map some of our features of interest to the space created by a CLIP model. Then, we discuss the rules and associations obtained and the importance of each feature considered. We apply this approach in two different data domains, clinical reports and film reviews, and compare the results obtained individually and when considering both. Finally, we discuss the limitations of this approach and how it could be further improved.
\end{abstract}


\section{Introduction}

Unstructured data representation is nowadays one of the most relevant sources of digital information \cite{jain2020review}. Whether it is video \cite{liu2020deep}, text \cite{Devlin2018Oct} or audio \cite{michelsanti2021overview}, most existing deep learning research is built with the aim of exploiting this information for all possible kinds of tasks and degrees of supervision \cite{ericsson2021well}. 

Most of the state-of-the-art deep learning models use the attention block, which has proven to scale particularly well with large amounts of data. One of these models is the Contrastive Language-Image Pre-training Model (CLIP) \cite{radford2021learning}, which encodes both text and image in the same vectorial space. CLIP embeddings have been used for zero-shot classification and image retrieval \cite{baldrati2022effective}, and have been successful in encoding texts and images from different domains \cite{thirunavukarasu2023large}.

However, CLIP embeddings are not always satisfactory for specific-domain applications and can present well-known biases \cite{agarwal2021evaluating, fumanal2021concept}. Since the target classes are specified natural language, CLIP performance can also be affected by how those classes were specified \cite{radford2021learning}. To mitigate those problems, it is necessary to further process the CLIP features or to fine-tune the model. Fine-tuning a CLIP model can be a very expensive procedure in terms of both data and computational cost, and different methods to alleviate this have been proposed in the literature \cite{10377476}. Additionally, any retraining performed on the original model can cause significant performance degradation in the model \cite{10377476}. Because of that, specific strategies have been developed for individual applications, such as art classification \cite{Conde_2021_CVPR} and clinical records analysis \cite{Lin2023PMCCLIPCL}. 

One of the most common procedures performed on CLIP features is visualization using T-SNE or PCA \cite{pareek2021data}. Many other procedures have been used to understand the embeddings obtained with these models, which is useful both for Explainable AI (XAI) and to exploit this knowledge in a larger machine learning pipeline \cite{joshi2021review}. Fuzzy rules can serve as useful tools for explainable AI \cite{9430516}, but they have not been used alongside CLIP embeddings for this purpose.

In this paper, we propose the use of fuzzy rules for XAI, using them to understand how the general feature embeddings are constructed by the CLIP models. Then, we perform feature extraction on the original data in order to look for variables that are semantically relevant. Finally, we map the values in those variables to the embedding space. In this way, we can study how the relevant variables in the problem affect the structure of the data in the embedding space.
We used data from two different data domains: reports from patients who suffered a stroke about their rehabilitation exercises and IMDB film reviews. We compare how the embeddings obtained in both data domains reflect the desired characteristics. 

The rest of the paper goes as follows: in Section \ref{sec:background}, we discuss some notions and literature about text processing, CLIP models and Fuzzy rule-based classification, which is important to understand the rest of the paper. In Section \ref{sec:methods}, we describe the data that we used and the fuzzy inference system that we applied. In Section \ref{sec:experimentation} we discuss the experiments performed and the results obtained. Finally, in Section \ref{sec:conclusions} we give our final conclusions for this work and state our future lines.

\section{Background} \label{sec:background}

\subsection{Text processing and sentiment Analysis}
Text processing refers to the manipulation, analysis, and extraction of information from raw textual data. Two fundamental steps in this process are tokenization and lemmatization. Tokenization is the initial step and consists of breaking down a piece of text into smaller units called tokens, which can be words, parts of words, or punctuation signs. Lemmatization, on the other hand, is a technique used to break down a word down to its root meaning, known as the lemma. Most tasks in natural language processing are computed after these two steps are performed \cite{chowdhary2020natural}.

Sentiment analysis is a very popular way of text processing. It focuses on the polarity of a text, and it can also study specific emotions (anger, happiness, sadness, etc.) and intentions (being interested or not interested). Techniques for sentiment analysis normally use lexicons with pre-fixed values for each word. These values can be used directly or further exploited using another algorithm \cite{bird2006nltk, hutto2014vader}. It is also possible to define custom features for classification \cite{bird2006nltk}. Deep learning models have achieved state-of-the-art in many natural language processing tasks \cite{lauriola2022introduction}. However, classical techniques are still used in many real-world applications, such as customer feedback analysis, social media monitoring, and brand reputation management \cite{kalyanathaya2019advances}.

\subsection{CLIP features}
CLIP feature embeddings were originally proposed for general zero-shot image classification problems \cite{radford2021learning}. The CLIP model can encode the textual representation of the different target classes and the image to classify into the same vectorial space. The predicted class is the one with a minimum cosine distance between the image and that vectorial space.

In addition to this initial application, additional research has tested the effectiveness of contrastive embeddings in other related domains, such as image captioning \cite{barraco2022unreasonable}, image retrieval \cite{baldrati2022effective} and video-to-text \cite{xu2021videoclip}. Besides using the CLIP model as it was originally trained in \cite{radford2021learning}, it is also possible to retrain or fine tune it to work in a particular domain. This is a popular line of research because the original CLIP model performs poorly in specific or niche tasks \cite{fumanal2021concept}, and because retraining a foundational model from scratch is expensive and requires large amounts of data, which might not even exist for some specific applications \cite{thirunavukarasu2023large}. 

Another possibility is to exploit the CLIP embedding space without fine tuning and use these features as a regularizing element \cite{garcia2019context, fumanal2023artxai} or as input for other machine learning pipelines \cite{Gao2021CLIPAdapterBV}. This approach is limited by the expressiveness of the original CLIP features but can be more efficient than fine-tuning the CLIP model if the additional machine learning methods are fast to compute. Existing research shows that even linear models can be good enough for domain adaptation \cite{Ouali_2023_ICCV}.

\subsection{Fuzzy rule-based classification}
The fuzzy rule-based classification consists of discriminating observations into different categories using rules that follow this structure \cite{kosko1986fuzzy}:
\begin{equation}
	\text{IF } \mathbf{x}_1 \text{ is } \mathbf{a}_{j1} \dots \mathbf{x}_n \text{ is } \mathbf{a}_{jn} \text{ THEN class } j \text{ for } j=1,\dots,C
\end{equation}.

Usually, a fuzzy rule-based classifier (FRBC) computes the degree of matching of an input with respect to all the rules and chooses the one that maximizes this value. Note that it is possible to use different criteria as the degree of matching. It is possible to take into account more than one rule and not only the one with maximum value \cite{wieczynski2022applying}. However, this complicates the interpretability of the system because the final decision now refers to more than one rule.    

\subsection{Related Works}
Fuzzy logic has emerged as a valuable tool for interpretable data classification in many different applications, which has laid the ground for significant contributions in the domain of XAI. In particular, a study \cite{indhuja2014fuzzy} introduced a fuzzy-based system designed for sentiment analysis, with a particular emphasis on product reviews. They specially study the effect of hedges and linguistic modifiers in the product review. Another noteworthy contribution in this context is \cite{liu2017fuzzy}, where a fuzzy inference system is compared to other machine learning methodologies, obtaining favourable results for the fuzzy system in both interpretability and accuracy. Fuzzy rules have also been used with multimodal data, \cite{vashishtha2020inferring}. In this study, the authors use fuzzy rules within a bigger machine learning pipeline so that the fuzzy classifier is used to make the final decision, using features obtained from both text and audio.

Some of the advantages of fuzzy logic have also been exploited in medical contexts because expert knowledge can be specified as rules and because those systems based on rules are also easier to understand for the final users of the machine learning system. In \cite{phuong2001fuzzy} the authors revise some of the ways in which the knowledge bases were constructed from medical knowledge.

\section{Methods} \label{sec:methods}

\subsection{Data used}

We have studied two data sets from different applications: a series of reports of adults about their mobility exercises after a stroke \cite{shafizadeh_2022}, and a well-known collection of IMDB film reviews. In the following, we refer to the first as the ``clinical'' dataset and the second one as the ``Film'' dataset \cite{imdb}.

\begin{itemize}
    \item The clinical dataset consists of the transcription of a series of interviews with older patients who have suffered a stroke. In the conversations, the individuals describe their opinions on the exercises that are part of their rehabilitation and the technology involved in them (online classes, wearable devices, apps, etc). This dataset consists of 51 comments from 33 different patients.
    \item The Film dataset consists of 50.000 reviews of popular movies on the IMDB website.
\end{itemize}

\subsection{Fuzzy Inference System} \label{sec:frbc}

\begin{figure}
	\subfloat[]{\includegraphics[width=.5\linewidth]{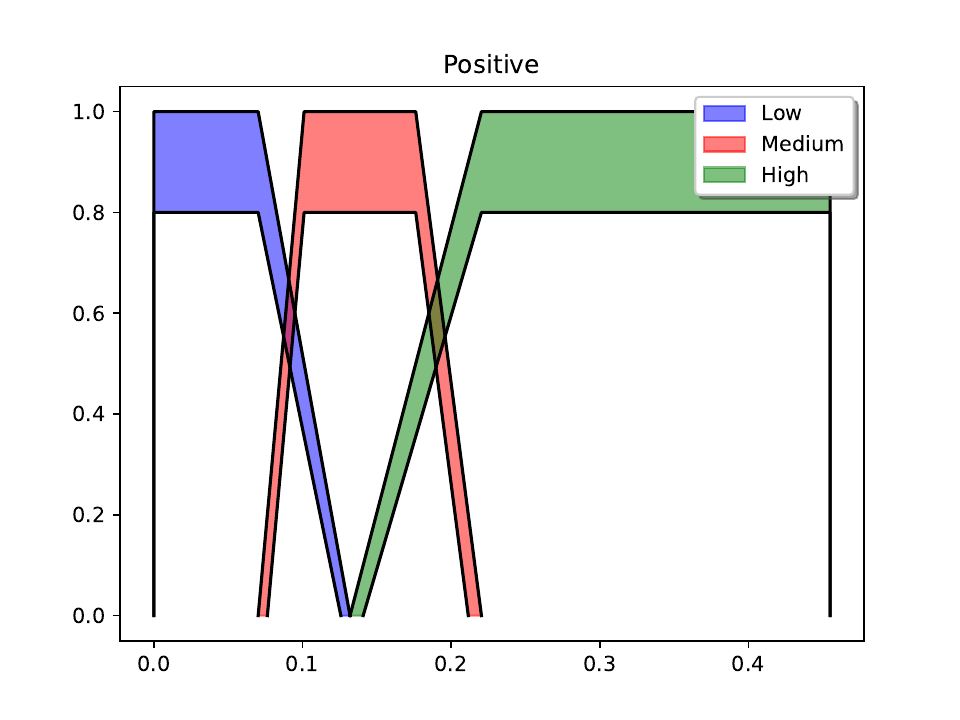}}
	\subfloat[]{\includegraphics[width=.5\linewidth]{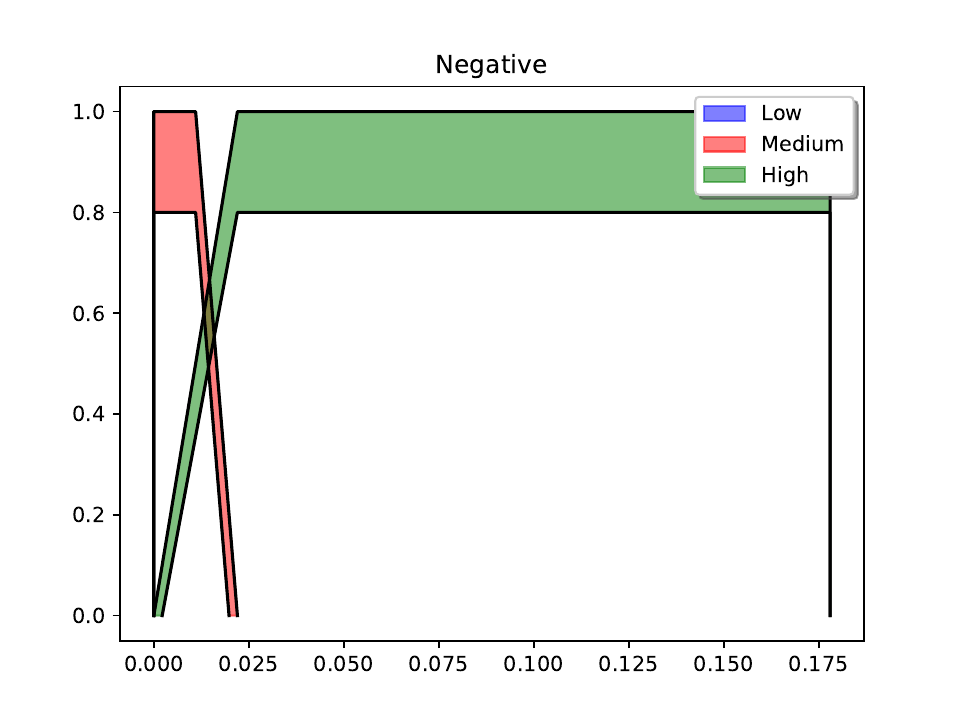}}\\
	\subfloat[]{\includegraphics[width=.5\linewidth]{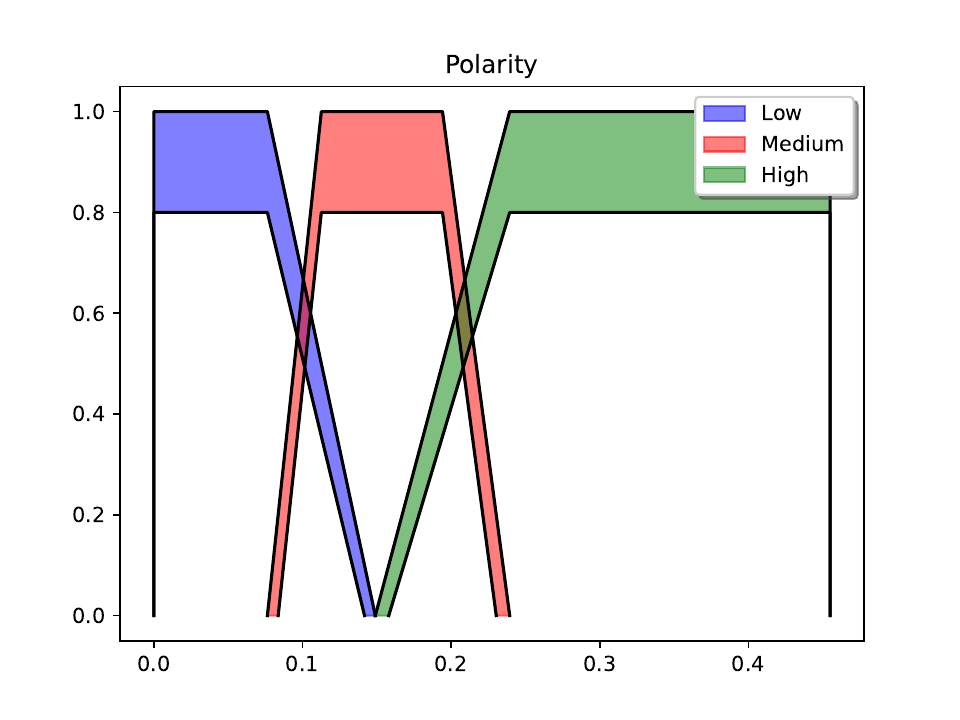}}
	\subfloat[]{\includegraphics[width=.5\linewidth]{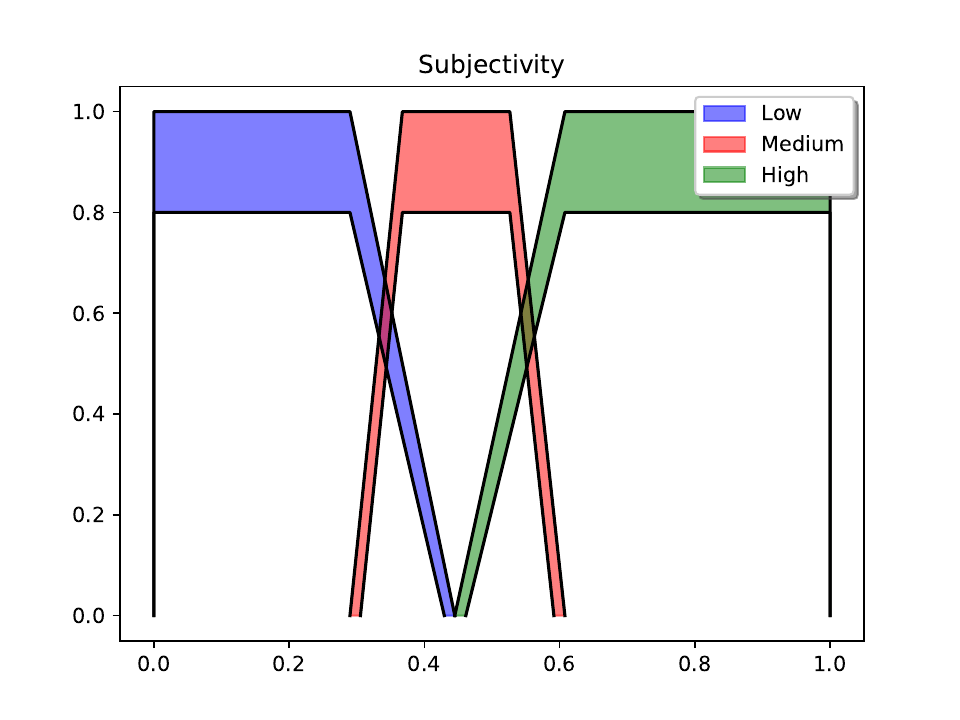}}
	\caption{Fuzzy partitions using interval-type 2 fuzzy sets for the considered variables in the clinical dataset.}
	\label{fig:partitions_t2}
\end{figure}

To obtain an FRBC that can be considered interpretable, we set $15$ as the maximum number of rules and $3$ as the maximum number of antecedents. However, the actual number of rules used will be reduced from that number using a quality metric. We also choose $3$ linguistic labels for the fuzzy partitions that are easily interpretable: low, medium, and high, which are the most intuitive to interpret. 

These partitions are computed empirically from the available data. The low partition has a maximum membership value from the minimum value to the 0.2 quantile and decreases until the 0.5 quantile. The medium partition begins at quantile 0.2, grows to its maximum value in the 0.5 quantile, and decreases until the 0.8 quantile. The high partition does the same thing for the quantiles 0.5, 0.8 and 1. When we used the interval-type 2 fuzzy sets, the lower membership had a maximum degree limit of $0.8$. (See them in Figure \ref{fig:partitions_t2}).
 
As a matching degree for each rule, we compute the dominance score, $ds_r$, of each rule $r$ \cite{kiani2022temporal, andreu2021explainable}. This metric is the product of two terms:
\begin{itemize}
	\item The a priori probability that a rule fires, $s_r$:
	\begin{equation} \label{eq:support}
		s_r = \frac{\sum_{\mathbf{x} \in 				Cons_r} w_r (\mathbf{x})}{|\mathbf{x} \in 				Cons_r|}.
	\end{equation}
	\item The rule firing strength compared to the rest, $c_r$:
	\begin{equation}
		c_r = \frac{\sum_{\mathbf{x} \in Cons_r} w_r (\mathbf{x})}{{\sum_{\mathbf{x} \in Cons_r }\sum_{r\prime=1}^{R}{w_r\prime(\mathbf{x})}}}.
	\end{equation}
	
\end{itemize}   

\noindent Where $w_r(\mathbf{x})$ is the firing strength of rule $r$, for the sample $\mathbf{x}$, $\mathbf{x} \in Cons_r$ is the set of observations whose ground-truth class corresponds to the rule class consequent and $R$ is the set of all rules. The firing strength is computed as the product of the truth degrees of all antecedents of the rule.

Finally, we compute the matching degree, $as_r(x)$, using $w_r(\mathbf{x})$ and $ds_r$:
\begin{equation}
	as_r(x) = w_r(\mathbf{x}) * ds_r.
\end{equation}

Each sample is classified according to the consequent class of the rule with the highest association degree for that sample. We also use the dominance score to prune those rules that did not reach a minimum value.

For our experimentation, we used the genetic algorithm to optimize the antecedents and consequents of each rule using the precomputed partitions. The metric to optimize is the Matthew correlation coefficient (MCC):
\begin{equation}
	MCC = \frac{(TP \times TN) - (FP \times FN)}{\sqrt{(TP + FP)(TP + FN)(TN + FP)(TN + FN)}}
\end{equation}
where TP is true positive, TN means true negative, FP is false positive, and FN is false negative.

We have also tried an additional loss function that adds two regularizing terms to the MCC. The regularizing terms penalize the size of the rule base based on the number of antecedents per rule ($l_1$) and the number of rules whose dominance score is bigger than a minimum threshold ($l_2)$:
\begin{equation}
    l_1 = \frac{1}{R} \sum_{i=1}^{R} \frac{nAnts(r_i)}{\text{maxAnts}}, 
\end{equation}

\begin{equation}
    l_2 = \frac{1}{R} \sum_{i=1}^{R} ds(r_i) > h, 
\end{equation}
\noindent where $R$ is the number of rules in the rule base, $h$ is the minimum threshold for a dominance score, maxAnts is the maximum number of antecedents allowed in a rule and $nAnts(r)$ is the number of antecedents of the rule $r$. 

The final loss is a convex combination between the MCC and the weighted sum of $l_1$ and $l_2$:

\begin{equation} \label{eq:full_loss}
    l = 0.95 * MCC + 0.05 * (0.50 * l_1 + 0.50 * l_2).
\end{equation}

\subsection{Exploiting CLIP embeddings using Fuzzy Rules}
\begin{figure*}
\centering
	\includegraphics[width=\linewidth]{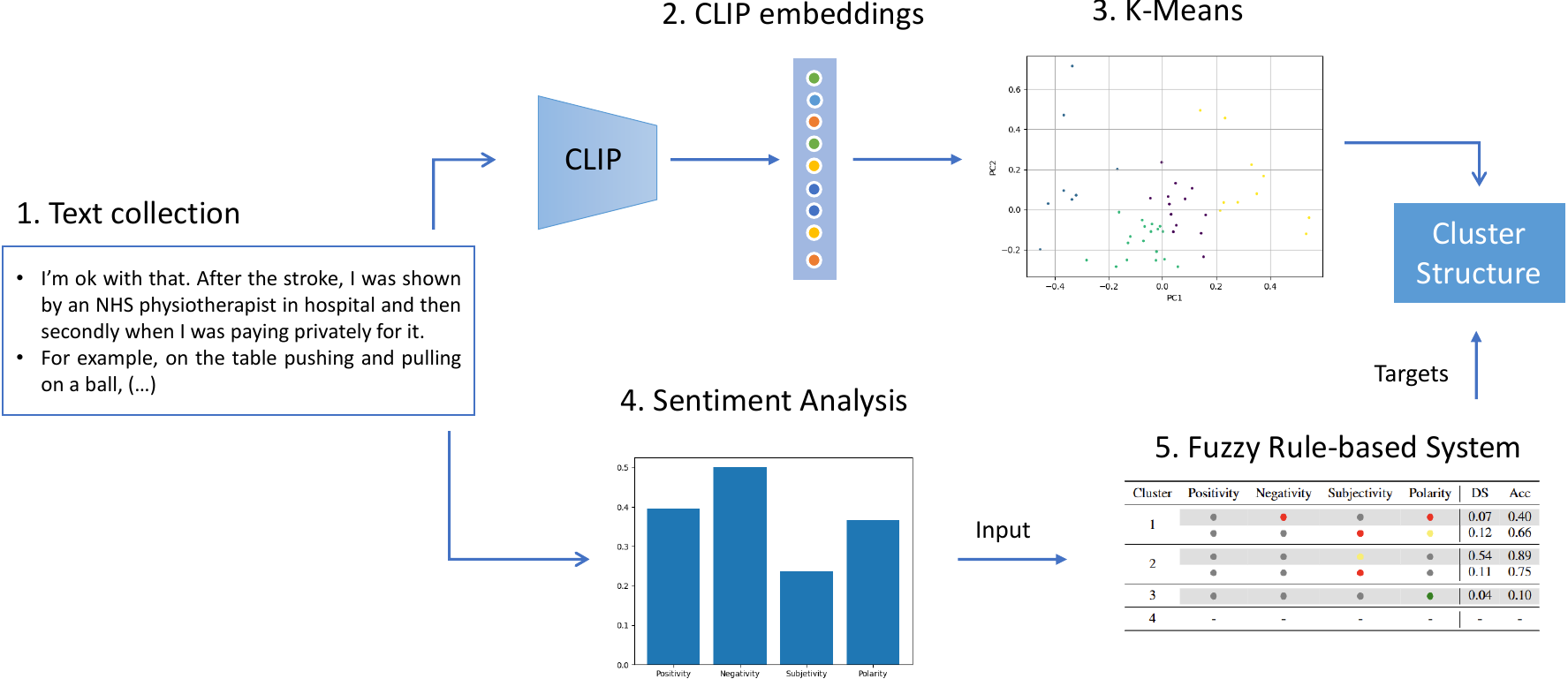}
	\caption{\textbf{Proposed methodology to exploit CLIP features and sentiment analysis using a FRBC.} 1. The collection of texts to study. 2. We compute the CLIP embeddings for all texts. 3. We use the K-Means clustering algorithm to obtain the structures formed in this space. 4. For each text, we extract the features of the sentiment analysis. 5. The FRBC: it takes as input for each text the sentiment analysis features, using the fuzzy partitions as shown in Figure \ref{fig:partitions_t2}. Then, it uses the clusters detected in Step 3 as targets. In this way, the FRBC maps our features of interest to the spacial regions where they were projected in their embedding space.}
	\label{fig:FRBC_m}
\end{figure*}

This process is divided into different steps:

\begin{enumerate}
    \item Obtaining the CLIP embeddings from the text: since CLIP embeddings are constructed using a maximum of $77$ characters, when the text to embed is longer than that, we split the text into parts of that size and compute the average of all the embeddings obtained for each split.
    \item Extracting the relevant features from the original texts. In this case, we are interested in the sentiment features from the texts: positivity, negativity, neutrality, and polarity \cite{bird2009natural}. In the case of patient reports, they represent the patient's opinion on their rehabilitation, and in the case of the film reviews, they represent the film's opinion.
    \item We look for a clustering structure in the embedded space. This is useful because, in particular cases, not all the dimensions are needed, and it is more reasonable to try to understand these groups than all the $512$ features in the embedding space. We construct these clusters using the k-means algorithm.
    \item We use the proposed FRBC in Section \ref{sec:frbc} to map the features to clusters in the CLIP space.
\end{enumerate}
We also report the accuracy and MCC of the FRBC, as well as the dominance score and accuracy of each individual rule.

Figure \ref{fig:FRBC_m} shows a visual representation of this process.

\section{Experimentation} \label{sec:experimentation}

\begin{figure}
\centering
	\includegraphics[width=.7\linewidth]{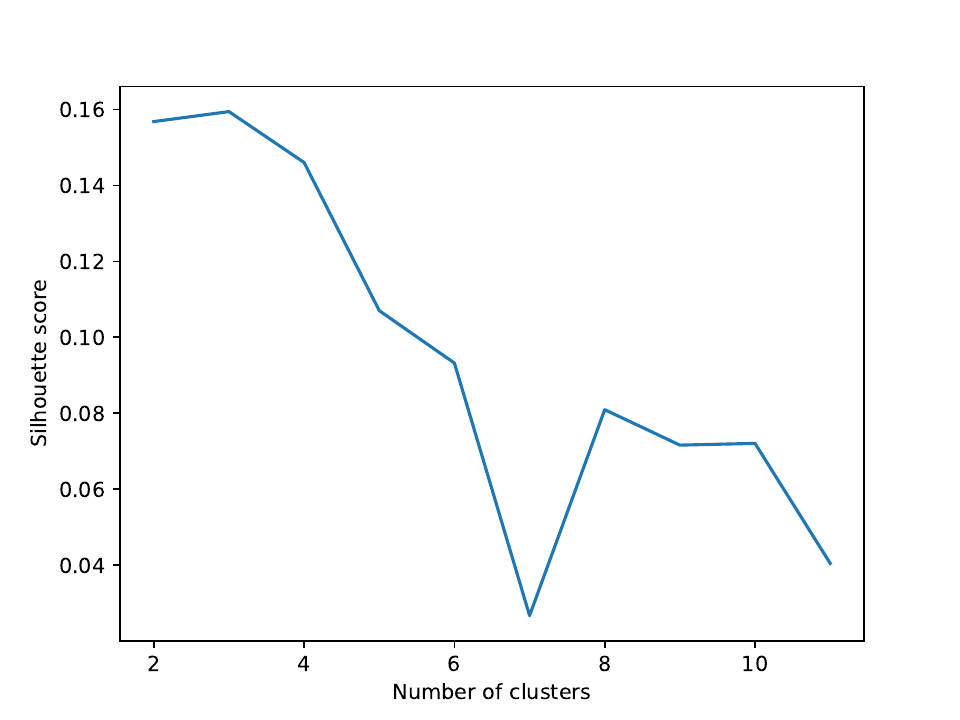}
	\caption{Silhouette index for K-Means clustering applied to the CLIP features obtained from all the patient's reports.}
	\label{fig:sil}
\end{figure}

\begin{figure}
\centering
	\includegraphics[width=.7\linewidth]{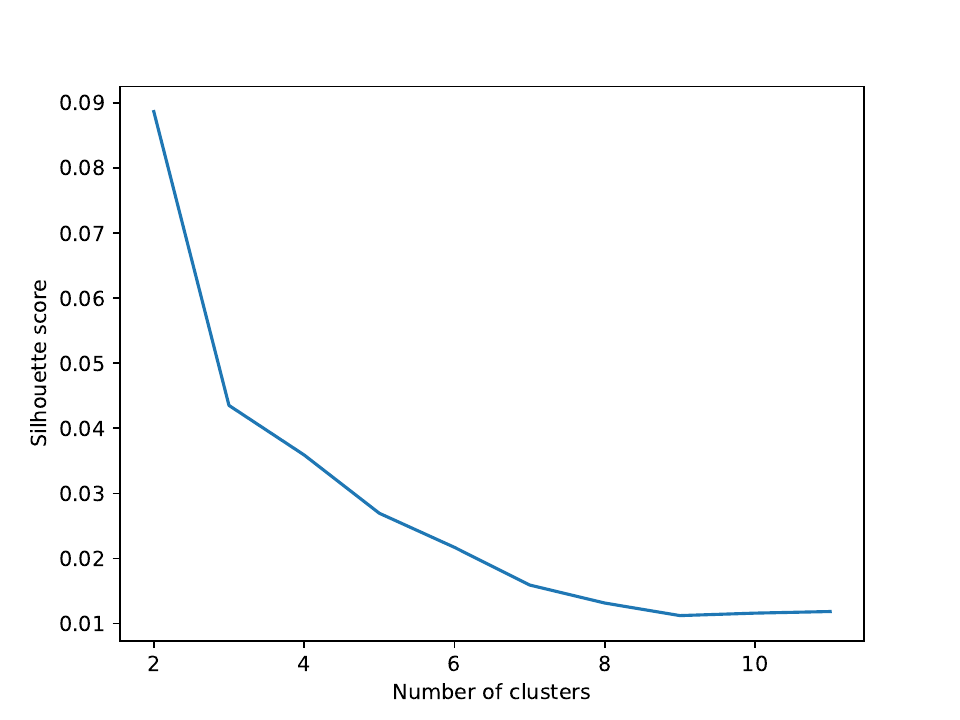}
	\caption{Silhouette index for K-Means clustering applied to the CLIP features obtained for the Film dataset.}
	\label{fig:sil_film}
\end{figure}

\begin{figure}
	\centering
    \subfloat[][MCC loss using IV fuzzy sets]{
	\adjustbox{max width=\linewidth, center}{
		\begin{tabular}{ccccc|cc}
			\toprule    
			Cluster &  Positivity & Negativity & Subjectivity & Polarity & DS & Acc \\
			\midrule
			\multirow{2}{*}{1}
			& \cellcolor{gray!25}\dc & \cellcolor{gray!25}\low & \cellcolor{gray!25}\dc & \cellcolor{gray!25}\low & \cellcolor{gray!25}0.07 & \cellcolor{gray!25}0.40\\
			& \dc & \dc & \low & \med & 0.12 & 0.66 \\
			\midrule
			\multirow{2}{*}{2} & \cellcolor{gray!25}\dc &  \cellcolor{gray!25}\dc & \cellcolor{gray!25}\med  & \cellcolor{gray!25}\dc  & \cellcolor{gray!25}0.54  &\cellcolor{gray!25}0.89\\
			& \dc & \dc & \low  & \dc & 0.11 & 0.75  \\
			\midrule
			3 & \cellcolor{gray!25}\dc & \cellcolor{gray!25}\dc & \cellcolor{gray!25}\dc & \cellcolor{gray!25}\hig &\cellcolor{gray!25}0.04 & \cellcolor{gray!25}0.10\\
            \midrule
            4 & - & - & - & - & - & - \\
			\bottomrule
			
	\end{tabular}}}

    \subfloat[][MCC loss using standard fuzzy sets]{
    	\adjustbox{max width=\linewidth, center}{
    		\begin{tabular}{ccccc|cc}
    			\toprule    
    			Cluster &  Positivity & Negativity & Subjectivity & Polarity & DS & Acc \\
    			\midrule
    			\multirow{1}{*}{1}
    			& \cellcolor{gray!25}\med & \cellcolor{gray!25}\low & \cellcolor{gray!25}\dc & \cellcolor{gray!25}\low & \cellcolor{gray!25}0.07 & \cellcolor{gray!25}0.75\\
    			\midrule
                \multirow{3}{*}{2}& \med & \dc & \dc & \hig & 0.17 & 0.85 \\
    			 & \cellcolor{gray!25}\dc &  \cellcolor{gray!25}\dc & \cellcolor{gray!25}\med  & \cellcolor{gray!25}\dc  & \cellcolor{gray!25} 0.42  &\cellcolor{gray!25}0.89\\
    			& \low & \low & \dc  & \dc & 0.06 & 1.00  \\
    			\midrule
    			\multirow{2}{*}{3} & \cellcolor{gray!25}\dc & \cellcolor{gray!25}\low & \cellcolor{gray!25}\dc & \cellcolor{gray!25}\med &\cellcolor{gray!25}0.06 & \cellcolor{gray!25}0.50\\
                & \dc & \hig & \dc & \dc & 0.02 & 1.00 \\
                \midrule
                4 & - & - & - & - & - & - \\
    			\bottomrule
    			
    	\end{tabular}}}
     
     \subfloat[][Loss in Eq. (\ref{eq:full_loss} using IV fuzzy sets]{
    	\adjustbox{max width=\linewidth, center}{
    		\begin{tabular}{ccccc|cc}
    			\toprule    
    			Cluster &  Positivity & Negativity & Subjectivity & Polarity & DS & Acc \\
    			\midrule
    			\multirow{1}{*}{1}
    			& \cellcolor{gray!25}\dc & \cellcolor{gray!25}\low & \cellcolor{gray!25}\low & \cellcolor{gray!25}\dc & \cellcolor{gray!25}0.06 & \cellcolor{gray!25}1.00\\
    			\midrule
                \multirow{3}{*}{2}& \low & \dc & \low & \dc & 0.16 & 0.75 \\
    			 & \cellcolor{gray!25}\low &  \cellcolor{gray!25}\dc & \cellcolor{gray!25}\hig  & \cellcolor{gray!25}\dc  & \cellcolor{gray!25} 0.06  &\cellcolor{gray!25}1.00\\
    			& \dc & \dc & \med  & \med & 0.33 & 0.94  \\
    			\midrule
    			3 & \cellcolor{gray!25}\dc & \cellcolor{gray!25}\dc & \cellcolor{gray!25}\dc & \cellcolor{gray!25}\hig &\cellcolor{gray!25}0.08 & \cellcolor{gray!25}0.16\\
                \midrule
                4 & \dc & \dc & \dc & \hig & 0.04 & 0.10 \\
    			\bottomrule
    			
    	\end{tabular}}}
	\caption[Results with rules obtained for the clinical dataset that map the sentiment metrics to the clusters in the CLIP space in the clinical dataset. (a) Using only the MCC as a loss. (b) Using the loss in Eq. (\ref{eq:full_loss}) using IV fuzzy sets]{Results with rules obtained for patient reports that map sentiment metrics to clusters in CLIP space. 
		 \low~$\rightarrow$~low, \med~$\rightarrow$~medium, \hig~$\rightarrow$~high, \dc~$\rightarrow$~irrelevant. DS stands for dominance score, and Acc. for the accuracy obtained by each rule in the samples where it fired.
		}
	\label{fig:table_medical_rules}
\end{figure}

\begin{figure}
	\centering

        \subfloat[][MCC loss using IV fuzzy sets] {
        \adjustbox{max width=\linewidth, center}{
        
    \begin{tabular}{ccccc|cc}
        \toprule
        Consequent & Positive & Negative & Polarity & Subjectivity & DS & Acc \\
        \midrule
        \multirow{5}{*}{1} & \cellcolor{gray!25}\med & \cellcolor{gray!25}\low & \cellcolor{gray!25}\low & \cellcolor{gray!25}\dc & \cellcolor{gray!25}0.02 & \cellcolor{gray!25}0.40 \\
 & \dc & \hig & \low & \dc & 0.06 & 0.70 \\
 & \cellcolor{gray!25}\low & \cellcolor{gray!25}\med & \cellcolor{gray!25}\low & \cellcolor{gray!25}\dc & \cellcolor{gray!25}0.10 & \cellcolor{gray!25}0.83 \\
 & \med & \dc & \med & \dc & 0.29 & 0.72 \\
 & \cellcolor{gray!25}\dc & \cellcolor{gray!25}\low & \cellcolor{gray!25}\low & \cellcolor{gray!25}\dc & \cellcolor{gray!25}0.03 & \cellcolor{gray!25}1.00 \\
        \midrule
        \multirow{2}{*}{2} & \dc & \hig & \dc & \dc & 0.07 & 0.39 \\
 & \cellcolor{gray!25}\low & \cellcolor{gray!25}\dc & \cellcolor{gray!25}\low & \cellcolor{gray!25}\dc & \cellcolor{gray!25}0.03 & \cellcolor{gray!25}0.24 \\
        \midrule
        \multirow{1}{*}{3} & \dc & \dc & \hig & \dc & 0.23 & 0.40 \\
        \bottomrule
\end{tabular}}} \\
    
    \subfloat[][MCC loss using standard fuzzy sets]{
    	\adjustbox{max width=\linewidth, center}{
        \begin{tabular}{ccccc|cc}
        \toprule
        Consequent & Positive & Negative & Polarity & Subjectivity & DS & Acc \\
        \midrule
        \multirow{2}{*}{1} & \cellcolor{gray!25}\low & \cellcolor{gray!25}\dc & \cellcolor{gray!25}\hig & \cellcolor{gray!25}\dc & \cellcolor{gray!25}0.02 & \cellcolor{gray!25}0.50 \\
 & \med & \dc & \dc & \hig & 0.08 & 0.37 \\
        \midrule
        \multirow{1}{*}{2} & \cellcolor{gray!25}\med & \cellcolor{gray!25}\hig & \cellcolor{gray!25}\hig & \cellcolor{gray!25}\dc & \cellcolor{gray!25}0.06 & \cellcolor{gray!25}0.44 \\
        \midrule
        \multirow{3}{*}{3} & \med & \med & \med & \dc & 0.15 & 0.56 \\
 & \cellcolor{gray!25}\low & \cellcolor{gray!25}\med & \cellcolor{gray!25}\low & \cellcolor{gray!25}\dc & \cellcolor{gray!25}0.14 & \cellcolor{gray!25}0.71 \\
 & \low & \dc & \med & \dc & 0.09 & 0.55 \\
        \bottomrule
\end{tabular}}}\\

          \subfloat[][Loss in Eq. (\ref{eq:full_loss} using IV fuzzy sets] {
        \adjustbox{max width=\linewidth, center}{
        \begin{tabular}{ccccc|cc}
        \toprule
        Consequent & Positive & Negative & Polarity & Subjectivity & DS & Acc \\
        \midrule
        \multirow{1}{*}{1} & \cellcolor{gray!25}\dc & \cellcolor{gray!25}\dc & \cellcolor{gray!25}\low & \cellcolor{gray!25}\low & \cellcolor{gray!25}0.01 & \cellcolor{gray!25}0.16 \\
        \midrule
        \multirow{3}{*}{2} & \dc & \dc & \low & \dc & 0.01 & 0.14 \\
 & \cellcolor{gray!25}\dc & \cellcolor{gray!25}\dc & \cellcolor{gray!25}\hig & \cellcolor{gray!25}\dc & \cellcolor{gray!25}0.15 & \cellcolor{gray!25}0.32 \\
 & \dc & \dc & \med & \dc & 0.03 & 0.19 \\
        \midrule
        \multirow{1}{*}{3} & \cellcolor{gray!25}\dc & \cellcolor{gray!25}\med & \cellcolor{gray!25}\dc & \cellcolor{gray!25}\dc & \cellcolor{gray!25}0.47 & \cellcolor{gray!25}0.83 \\
        \bottomrule
\end{tabular}}}

	\caption[  Rules obtained for the patient reports that map the sentiment metrics to the clusters in the CLIP space in the Film dataset. (a) Using only the MCC as a loss. (b) Using the loss in Eq. (\ref{eq:full_loss})]{Results with rules obtained for the film reviews that map the sentiment metrics to the clusters in the CLIP space. 
		 \low~$\rightarrow$~low, \med~$\rightarrow$~medium, \hig~$\rightarrow$~high, \dc~$\rightarrow$~irrelevant. DS stands for dominance score and Acc. for the accuracy obtained by each rule in the samples where it fired.
		}
	\label{fig:table_film_rules}
\end{figure}

\begin{table}[h]
    \centering
    \adjustbox{max width=\linewidth, center}{
    \begin{tabular}{cccccc}
    \toprule
    Dataset & Loss & Fuzzy Set & Accuracy & Rules & Antecedents \\
    \midrule
    \multirow{4}{*}{Clinical} & Eq. \ref{eq:full_loss}  & T1 & $0.64 \pm 0.09$ & $4.60\pm1.01 $& $11.40\pm2.57$\\
                              & Eq. \ref{eq:full_loss}  & T2 & $0.64 \pm 0.07$ & $4.40\pm1.01$& $10.00\pm2.68$\\
                              & MCC                     & T1 & $0.66 \pm 0.08$ & $5.50\pm0.92$& $13.00\pm2.14$\\
                              & MCC                     & T2 & $0.70 \pm 0.06$ & $5.80\pm1.72$ & $14.10\pm3.75$ \\
    \midrule                          
    \multirow{4}{*}{Film}       & Eq. \ref{eq:full_loss}  & T1  &  $0.46\pm0.96$ & $4.83\pm1.06$ & $15.54\pm0.10$\\
                                & Eq. \ref{eq:full_loss}  & T2  & $0.47\pm0.01$  &  $4.10\pm1.00$& $17.60\pm3.44$\\
                                & MCC & T1 &  $0.44\pm0.01$ & $7.00\pm1.00$ & $11.58\pm1.02$\\
                                & MCC & T2 & $0.46\pm0.22$ & $6.91\pm1.05$ & $14.00\pm2.53$\\
         \bottomrule
    \end{tabular}}
    \caption{Performance report of the different FRBC tested according to the different data configurations, losses and fuzzy sets tested. Results are the average for $30$ trials.}
    \label{tab:performance_report}
\end{table}
For our experiments, we used a CLIP model to embed all the texts in the dataset in the same feature space. Then, we used a K-Means to look for clusters in this representation. To determine the optimal number of clusters, we performed a Silhouette index analysis for their various values. We found that both datasets $3$ and $4$ could be a suitable cluster number for these data because the index decreases sharply with more groups (Figures \ref{fig:sil} and \ref{fig:sil_film}). We also considered other clustering algorithms, but the results were very similar.

Once we have computed the clusters, we fit the FRBC as described in Section \ref{sec:frbc} using the genetic algorithm. We optimize it for 300 generations using 30 subjects per generation. In order to obtain statistical significance of our results, we executed the results $30$ times.

Figure \ref{fig:table_medical_rules}a shows an FRBC for clinical reports using only MCC as a loss. This rule base obtained an accuracy of $0.63$ and a MCC of $0.47$. In this case, we obtained a total of $5$ rules, in which the attribute ``Subjectivity'' was the one that appeared in most rules, and the average number of antecedents is $2.5$. With standard fuzzy sets, the best FRBC leads to an MCC of $0.56$ and an accuracy of $0.81$ (FRBC shown in Figure \ref{fig:table_medical_rules}b). Finally, we show the results when we use the loss displayed in Eq. (\ref{eq:full_loss}). In this case, we obtain an accuracy of $0.69$ and an MCC of $0.33$. This FRBC learned $6$ rules, in which the average number of antecedents is $1.66$.

When repeating the same experiments with the Film dataset, we obtained an accuracy of $0.36$ and $0.09$ using standard fuzzy sets and MCC loss. We obtained similar results ($0.40$ of accuracy and $0.10$ of MCC) using a gradient-boosting classifier, which is considered the state-of-the-art classification in tabular data \cite{natekin2013gradient}. This dataset is harder as the variance in vocabulary between the samples is higher. Thus, the structure formed in this space is much less affected by our features of interest. What we did, in this case, was study a local interaction by choosing one random sample and then choosing the $1000$ closest samples to this in the CLIP space using Euclidean distance (cosine distance yields similar results). This approach is similar to other explainable AI studies that consider local interactions between variables \cite{lundberg2020local}.


Figure \ref{fig:table_film_rules} shows the rule bases for this case. The rulebases obtained were of a similar size compared to those obtained in the clinical dataset. We found that most of the time, one cluster is clearly identified, but the rest require more complex patterns. We can also observe that most rules largely ignored the subjectivity feature and how polarity was considered in most rules.

As shown in Table \ref{tab:performance_report}, the best configuration was using standard fuzzy sets and the loss function in Eq. \ref{eq:full_loss} we obtained an average accuracy of $0.49$. We can also observe that while the number of rules was reduced with respect to the MCC loss, it was not the case for the number of antecedents.


\subsection{Discussion}
Our results show the effectiveness of using fuzzy rules and sentiment analysis features to understand the structure formed in a particular dataset using a CLIP model. The CLIP model embeds the text, taking into account all the characteristics of the text, and using a fixed number of dimensions, which is an advantage over vector representations. This FRBC was significantly more successful in the case of the clinical dataset than in the IMDB data set, where the rule sets obtained had more rules and were less effective. We believe this is because the clinical dataset has a much lower number of samples, and the variability of the text contents was also more limited. The results in both cases seem to indicate that additional variables could benefit the rulebase performance. In order to reduce the number of additional variables, we could use feature engineering to artificially elaborate those that are most beneficial to the FRBC. The joint use of such variables and interpretable variables for XAI is an active research topic \cite{lundberg2020local}.

We tried two different loss functions to optimize: one that only took care of the classification performance, and another that also took into account the rulebase size. The effects of the second loss in the optimization process resulted in smaller rule bases, but the precision was also degraded. This trade-off between accuracy and explainability is a well-known issue in FRBC \cite{moral2021explainable}. We also found that the type-2 interval fuzzy sets obtained better results than those using standard fuzzy sets using the MCC loss. We can attribute this increase in performance to the more flexible membership functions and the more efficient way in which these fuzzy sets partition the input space \cite{mendel2017uncertain}. However, when we use the size-aware loss, this positive effect disappears.

Finally, we also found that in some rules, the dominance score and the accuracy can be very different. This was particularly notorious in the clinical dataset rulebases (See Figure \ref{fig:table_medical_rules}). The dominance score is a measure of how good a rule is based on the generality of its pattern and its uniqueness with respect to other rules in the FRBC. The reason for this disagreement between accuracy and dominance scores is that these rules have captured local patterns in the clusters. It is possible to build global explanations from them \cite{lundberg2020local}, but it would require building a FRBC specifically to find them.

\section{Conclusions and Future Lines} \label{sec:conclusions}
In this paper, we have presented a method to interpret the embedding space of a CLIP model using fuzzy rules. First, we have explained the features extracted from the original text using sentiment analysis. Then, in order to look for the relevant structures in the embedding space for each dataset, we used k-means. Finally, we mapped the degree of expression of these features to the clusters in the CLIP dataset. We show the effects of using different fuzzy sets and loss functions in the FRBC size. We applied this methodology in a dataset of patient transcripts discussing their experience with motor rehabilitation exercises and in an IMDB film review dataset. Finally, we also compared the differences in the FRBC when using each dataset individually.

Our results show that when the variability among the themes of the texts is narrow enough, this mapping can capture the relationship between the structures formed by the CLIP embedding and the desired features. However, this mapping is not perfect, and additional information would be necessary to obtain a more accurate mapping. The interval-type 2 fuzzy sets obtained better results than the standard fuzzy sets. We also found that penalizing the number of rules and rule sizes resulted in significant penalization in the final MCC of the FRBC. Thus, the trade-off between performance and explainability is still to be explored in future work.

Future lines should explore the embedding spaces of other clinical data in order to obtain representations that can capture the state of one patient and make its comparison with others possible. We also intend to use the FRBC alongside other local and model-agnostic explainability methods.

 \section{Acknowledgment}
 Javier Fumanal-Idocin research has been supported by the European Union under a Marie Sklodowska-Curie YUFE4 postdoc action.

\bibliographystyle{IEEEtran}
\bibliography{wcci}

\end{document}